\documentclass[11pt,eqs]{article}
\usepackage{latexsym}

\title{Charge dependent relation between the masses of different generations
and Neutrino masses
\thanks{Work supported in part by the Serbian Ministry of Science and
Technological Development, under contract No. 171031.}}
\author{B. Sazdovi\'c
\thanks{e-mail address: sazdovic@ipb.ac.rs}\\
{\it Institute of Physics,}\\
{\it University of Belgrade,}\\
{\it 11001 Belgrade, P.O.Box 57, Serbia}}

\begin{document}
\maketitle

\begin{abstract}
Despite the enormous achievements, the Standard model of
Particle physics can not be consider as complete theory of
fundamental interactions. Among other things, it can not describe
the gravitational interaction and it depends on 19 
parameters. The Standard model includes 12 fermions (matter
elementary particles with spin $\frac{1}{2}$) which are divided in
three generations, groups with same interactions but
different masses. Each generation can be classified into two
leptons (with electric charges $Q=-1$, electron-like and $Q=0$,
neutrino) and two quarks (with electric charges $Q=-\frac{1}{3}$,
down-type and $Q=\frac{2}{3}$, up-type). However, the
understanding of the relationship between generations and ratio of
masses of different generations are unknown. Here we show that
there exists the simple relation between masses of different
generations which depend only on the electric charges for $Q=-1,\,
\, Q=-\frac{1}{3}$ and $Q=\frac{2}{3}$. It is in pretty good
agreement with experimental data. Assuming that the same relation
valid for $Q=0$, we are able to calculate neutrino masses.
Therefore, our results could pave the way for further
investigations beyond Standard model.
\end{abstract}

\section{The basic relation}

Although the Standard model successfully explains many experimental results, it can not explain
origin of quarks and leptons generations and relation between
their masses.
In the present article we offer the simple relation which connect
the ratio of masses of different generations with their
charges.

It is well known that there are three generations of quarks and leptons in Nature.
They appear in equal numbers and with the
same interactions in every generation. Only the masses vary, but with significant
difference. From neutrino oscillations it is known
that neutrinos have masses, but experiments gives only an upper limit.

\begin{table}[hbt]
\caption{Masses and charges of quarks and leptons
(the $t$ quark mass is in GeV, while all other masses are in Mev)\label{t1}}
\vspace*{3mm}
\begin{tabular}{|c||c||c||c||c||c|}\hline
      &   $Q=-1$      &  & $Q=-\frac{1}{3}$ & &   $Q= \frac{2}{3}$ \\      \hline \hline
$m_{\tau}$  &  1 776.82 $\pm$ 0.16     & $m_{ b}$ & ($\overline{MS}$): \,\, 4 180  $\pm$ 30   & $m_{t}$  &   173.21  $\pm$  0.51  $\pm$  0.71     \\
  &     &   &  (1S): \,\, 4 660  $\pm$ 30  &   & $(\overline{MS}): \,\,  160^{+5}_{-4}$     \\
  &     &   &                          &   & $    176.7^{+4.0}_{-3.4}$     \\ \hline \hline
$m_{\mu}$  &    105.6583715 $\pm$ 0.0000035     &  $m_{s}$ & 95  $\pm$ 5    & $m_{c}$  &  1 275  $\pm$  25  \\ \hline \hline
$m_{e}$  &   0.510998928  $\pm$  0.000000011      &  $m_{d}$ &  4.8 ${}^{+ 0.5}_{-0.3}$     & $m_{u}$  &   2.3 ${}^{+ 0.7}_{-0.5}$    \\ \hline \hline
\end{tabular}
\end{table}

Let us start with the  Table 1 for three generation  of quark and
lepton masses taken from Review of Particle Physics
\cite{RPP}.
We propose a simple relation  which connect dimensionless mass
dependent expression (on the left side) with simple charge dependent
expression (on the right side)
\begin{equation}\label{1}
\frac{M_2^2}{M_1 \, M_3} = e^{\frac{5}{2}Q(q-l)} \,  .
\end{equation}
Here $M_1=\{m_e,m_d,m_u\}, \,\, M_2=\{m_\mu,m_s,m_c\}$ and
$M_3=\{m_\tau,m_b,m_t \}$ are masses from first, second and third
generations, $Q$ is a electric charge while $q$ and $l$ are the
quark and lepton numbers.
Note that right hand side depend on the square of quantum
numbers and consequently does not depend on the  particle
antiparticle replacement.

\begin{table}[hbt]
\caption{Theoretical values \label{t2}} \vspace*{3mm}
\begin{tabular}{|c||c||c|}\hline
         $Q=-1, l=1, q=0  $      &   $Q=-\frac{1}{3},l=0, q=1$ &    $Q= \frac{2}{3}, l=0, q=1$ \\      \hline \hline
$e^{\frac{5}{2}}$ =12,18249   & $e^{-\frac{5}{6}}$ =0,434598     &
$e^{\frac{5}{3}}$ =5,29449  \\ \hline \hline
\end{tabular}
\end{table}

The right hand side of (\ref{1}) is easy to calculate and for
different values of charges we obtain Table 2. For the left hand
side of (\ref{1}) one should use experimental data. They are well
known for charge leptons (electron, muon and tau), they are
roughly estimated for quarks and they are not known for neutrinos.

\subsection{Lepton sector}

In the lepton sector we have
\begin{equation}\label{ls}
\frac{m_{\mu}^2}{m_{e} \, m_{\tau}}= 12,2939  \, .
\end{equation}
Because we take into account only electro-magnetic interaction, it
is in pretty good agreement with theoretical value.

\subsection{Quark sector}

The masses in the Table 1 are taken from Ref.\cite{RPP}.
The masses of the first generation $m_u \, , m_d$ and mass of second
generation $m_s$ are estimates of so-called "current-quark masses" in a mass-independent
subtraction scheme such as $\overline{MS}$ at a scale $\mu \approx 2 \, \,
GeV$. The second generation $m_c$ and third generation $m_b$
masses are "running" masses in the $\overline{MS}$ scheme. For the
b-quark the $1S$ mass is also quoted.
For the third generation $m_t$ quark mass the first value is from
direct measurements, the second one is in $\overline{MS}$ scheme from
cross-section measurements while the third one is pole from
cross-section measurements.

\begin{table}[ht]
\caption{Upper bounds, estimated values and lower bounds of $\frac{M_2^2}{M_1 \, M_3}$ for quarks \label{t3}}
\vspace*{3mm}
\begin{tabular}{|c||c||c||c||c||c|}\hline
 &          $Q=-\frac{1}{3}$ &       &  $Q= \frac{2}{3}$ \\      \hline \hline
                  &  Max  &  0,5355   &                    & Max   &  5,4590  \\
  $\overline{MS}$ &       &  0,4498   & Direct measurements  &       &  4,0806 \\
                  &  Min  &  0,3630   &                       &  Min   &  2,9859  \\ \hline \hline
                  &  Max  &  0,4800   &                    &  Max   &  6,2593  \\
    1S            &       &  0,4035   &  $\overline{MS}$   &        &  4,4175  \\
                  &  Min  &  0,3259   &                       &  Min   &  3,1566  \\ \hline \hline
                  &   Max &           &                      &  Max   &  5,4177  \\
                  &       &           &   Pole from cross-section  &          &  4,0000  \\
                  &  Min  &           &                        &  Min   &  2,8823  \\ \hline \hline
\end{tabular}
\end{table}

The upper bound, the estimated value and the lower bound
for all cases of quark masses are calculated in the Table 3.
We denote Max $\equiv \frac{(max M_2)^2}{min M_1 \,\, min M_3}$
and Min $\equiv \frac{(min M_2)^2}{max M_1 \,\, max M_3}$.
In all cases the indeterminacy is not small, but the theoretical
values from Table 2 lies within allowed intervals.

\section{Neutrino masses}

Let us suppose that the proposed expression (\ref{1}) valid for neutrinos,
also. Then for $Q=0$ we have
\begin{equation}\label{2}
m_{\nu_\mu}^2 = m_{\nu_e} m_{\nu_\tau} \, .
\end{equation}
In the neutrino sector the only known experimental values  are data based on the 3-neutrino
mixing scheme \cite{RPP}
\begin{equation}\label{3a}
\Delta m_{21}^2 \equiv m_{\nu_\mu}^2 - m_{\nu_e}^2 =(7.53 \pm 0.18) \times 10^{-5} eV^2
\, , \quad \Delta m_{32}^2 \equiv m_{\nu_\tau}^2 - m_{\nu_\mu}^2 =(2.44 \pm 0.06) \times
10^{-3} eV^2  \,  ,
\end{equation}
for normal mass hierarchy ($m_{\nu_e} < m_{\nu_\mu} < m_{\nu_\tau}$) and
\begin{equation}\label{3b}
\Delta m_{32}^2 \equiv m_{\nu_\mu}^2 - m_{\nu_\tau}^2 =(2.52 \pm 0.07) \times
10^{-3} eV^2  \,  .
\end{equation}
for inverted mass hierarchy ($m_{\nu_\tau} < m_{\nu_e} <
m_{\nu_\mu}$). So, the sign of $\Delta m_{32}^2$ is not known
experimentally. In our case, from the relation (\ref{2}) we can
conclude that it has the same sign as $\Delta m_{21}^2$.
Therefore, the normal mass hierarchy appears.

Using expression (\ref{2}) we have
\begin{equation}\label{4}
m_{\nu_e} =  {\Delta m_{21}^2 \over {\sqrt{\Delta m_{32}^2 - \Delta m_{21}^2}} }\, , \quad
m_{\nu_\mu} = \sqrt{ {\Delta m_{21}^2 \, \Delta m_{32}^2 \over {\Delta m_{32}^2 - \Delta m_{21}^2}}
}\,  , \quad
m_{\nu_\tau} =  {\Delta m_{32}^2 \over {\sqrt{\Delta m_{32}^2 - \Delta m_{21}^2}} }
\,  .
\end{equation}
The explicit values are present in Table 4, which is supplement of
Table 1 for $Q=0$.

\begin{table}[hbt]
\caption{Neutrino masses  \label{t4}}
\vspace*{3mm}
\begin{tabular}{|c||c|}\hline
          &    $Q=0$       \\      \hline \hline
$m_{\nu_\tau}$   &  50.18   $\times 10^{-3} eV$       \\ \hline \hline
$m_{\nu_\mu}$   &   8.81   $\times 10^{-3} eV$         \\ \hline \hline
$m_{\nu_e}$     &   1.55   $\times 10^{-3} eV$       \\ \hline \hline
\end{tabular}
\end{table}

All results for neutrino masses are in agreement with limitation
from tritium decay $m_\nu < 2 eV$, \cite{RPP}.
Consequently, if the relation (\ref{1}) is true for neutrinos, we
are able to calculate their masses.

\section{Conclusion}

The central part of the present article is the expression
(\ref{1}). It shoes that the dimensionless mass dependent expression
of different generations depend only on electrical charge.
We are not able to prove this expression theoretically, but we verified
it in the cases of charge leptons and quarks, using experimental data from Ref.\cite{RPP}.
We can conclude that the relation (\ref{1}) is in pretty good agreement with experimental
data.

The important prediction follows  from the hypothesis that the expression
(\ref{1}) valid for uncharged leptons--neutrinos, also. In that
case we are able to calculate neutrino masses (see Table 4).

The only parameter in equation (\ref{1}) is $\frac{5}{2}$. We do
not inclined to explain the origin of this number, but those who
is convinced of the correctness of the string theory can just argue that this is a quotient of
number of critical dimensions in superstring theory $D=10$ and the number of
non-compactified dimensions $d=4$.

Consequently, if our main relation is true, there are 6 instead of
9 independent masses and 16 instead of 19 independent parameters
in the Standard model. In addition, we predicted neutrino masses
which are not the part of the Standard model.
We hope that new relation will help  better understood the problem of
generations in particle physics. Because the neutrino masses have
not yet been measured we expect that future experiments will confirm
our prediction.

\end{document}